\documentclass[preprint, 12pt]{elsarticle}
\usepackage[utf8]{inputenc}
\usepackage{graphicx}
\usepackage{subfigure}
\usepackage{color}
\usepackage{natbib}
\usepackage{textgreek}
\usepackage{upgreek}
\usepackage{multirow}
\usepackage{amsmath}
\usepackage{amssymb}
\usepackage{url}
\usepackage{comment}
\title{A Civilian Astronomer's Guide to UAP Research}

\address[UA]{Nordita, KTH Royal Institute of Technology and Stockholm University, Roslagstullsbacken 23, SE-106 91 Stockholm, Sweden}
\address[PX]{Department of Physics and Astronomy, Texas A\&M University, 4242 TAMU, College Station, Texas 77843, USA}

\author[UA]{Beatriz Villarroel*}
\cortext[]{Corresponding author. \textit{E-mail address}: \\ beatriz.villarroel@su.se (Beatriz Villarroel).}
\author[PX]{Kevin Krisciunas}

\begin{document}

\begin{abstract}

Unidentified Anomalous Phenomena (UAP) have historically been stigmatized and regarded as pseudoscience due to a general lack of robust evidence. Recently, however, the subject has gained interest among astronomers and the military. This review explores how astronomers can enhance our understanding of these enigmatic phenomena by focusing on empirical tests of specific hypotheses (e.g. the hypothesis of extraterrestrial visitations) rather than solely collecting and categorizing data. We compare the investigation of UAP to the process of calibration and interpretations of astronomical discoveries and propose a toy model involving a network of neuro-interface extraterrestrial probes to model exotic UAP. This model aids in predicting probe signatures and behaviour, improving detection methods, and addressing ethical concerns in UAP research.

\end{abstract}
\maketitle

\section{Introduction}

The study of Unidentified Aerial Phenomena, also known as Unidentified Anomalous Phenomena (UAP) is a new multidisciplinary field of research, originating from the topic of Unidentified Flying Objects (UFOs). The UFO field has historically been stigmatized \citep{Barkun2015}, and the proposed rebranding of the phenomenon was encouraged as a means of encouraging a scientific approach to the mystery, not least by the United States Department of Defense \citep{Pasulka2019,Stahlman2024}. The topic has further been discussed openly in the US Congress see e.g. news item by Romo \& Chapell (2023) \cite{Romo2023} due to its importance for airspace safety, national security, and recently also in a meeting at the European Parliament see e.g. news item by Fleischer (2024) \citep{Fleischer2024}. But before ``UAP'' and ``UFO'', the phenomenon was dubbed ``flying saucers'' in the media, referring to pilot Kenneth Arnold's observations of flat saucer-shaped disks in the air in 1947 \citep{Arnold1947,Hall1998}. In this way, the description of the phenomenon changed from an airborne phenomenon with distinct artificial features that were seen by eyewitnesses on multiple occasions and sometimes even photographed, to become a category of all objects in the air, space, or water that do not match known human-made or natural phenomena. Some of them appear to show transmedium capabilities \citep{Gallaudet2023}, inconsistent with any known human objects.

The change of nomenclature from ``flying saucer'' to ``UAP'' is intended to reduce the stigma and ridicule related to the topic of ``little green men'' \citep{Barkun2015}. However, as a side effect, it transformed a phenomenon with clear observational signatures into a ``junk drawer'' of objects with no defined signatures other than that their properties match no known natural or human-made objects, see Section \ref{sec:taxonomy}.

Due to the same stigma, the topic has been poorly researched in academic environments until recent years. As the US Navy released some videos of UAP in 2020 \citep{Strauss2020} and the Pentagon further published a UAP report in 2021 \citep{ODNI2021}, the trend shifted, resulting in an increase of academic interest. See for example the \textit{Galileo} project\footnote{https://projects.iq.harvard.edu/galileo/home} \citep{Galileo,Watters2023}.

The stigma can be seen as a societal effect fueled by a number of factors, some within our control and some related to national security interests \citep[see e.g. news item by][]{Eghigian}. Among scientists, one of the dominant factors has been the poor data quality of observations; many UFO observations have been based on fuzzy images.  Those based on sharp images of flying saucers are often accused of being fakes, see e.g. the Swedish newspaper article by Svahn (2024) \cite{Svahn2024}. Additionally, there is a deeply rooted fear of sensational performances and potential consequences. This dates back to 1938 when a radio play ``The War of the Worlds'' was broadcast by Orson Welles. It was mistaken by some as a real alien invasion. In contradiction to the legend, there was no mass panic \citep{Pooley}.  Nevertheless, the legend  about the mass panic enhanced the stigma even further.

The stigma is particularly strong regarding the hypothesis that UAP might be artificial objects from other civilizations. This contrasts with ongoing searches for technosignatures at interstellar distances, such as radio waves. The latter has been ongoing since the 1960s, with no results, even after \textit{Breakthrough Listen} began the most systematic radio searches to date. Searches within our Solar System are colloquially associated with fringe science, and many scientists instinctively react negatively to the mere suggestion that another civilization might have visited our Solar System or Earth\footnote{Some Searches for Extraterrestrial Intelligence (SETI) conferences exclude the topic of near-Earth searches  for artifacts and UAP from their program.}. As a result of this bias, astronomers tend to respond more emotionally \footnote{See e.g. this prose, ``Avi Loeb and ‘Oumuamua. Why the controversy?'', https://earthsky.org/human-world/avi-loeb-oumuamua-controversy/} to suggestions that ‘Oumuamua could be a ET spaceship \citep{Bialy} than to suggestions that some stars could be Dyson sphere candidates \citep{Suazo2024}. Yet humans are far more likely to be capable of building a spaceship or exploratory probe than a Dyson sphere — our own Voyager and Pioneer probes have already left the Solar System. Even in the early works of Carl Sagan \citep[see, for example Sagan (1963)][]{Sagan1963}
he estimated that Earth may have been visited $10^{4}$ times by civilizations travelling at near-light speeds \footnote{A few years later, the UFO friendly Carl Sagan underwent a transition and became a UFO skeptic.}. We can only speculate on the advances in physics that a civilization a million years older than ours might have made, some of which could encourage interstellar travel. Under such circumstances, the question of the possible non-human intelligence (NHI) origin of UAP becomes relevant.

Much of the support for the UAP phenomenon comes from eyewitness accounts. Detailed eyewitness reports of UAP are valuable and can help discern between a number of explanations for UAP sightings, particularly when experienced observers such as advanced amateur astronomers or professional pilots, who know the sky better than other groups, provide testimony. Report centers exist worldwide, and organizations such as MUFON\footnote{https://mufon.com} and GEIPAN\footnote{https://www.cnes-geipan.fr/} have collected tens of thousands of eyewitness reports over the past decades. However, we must remind ourselves that humans do not see with their eyes but with their brains, which are uncalibrated sensors for sight, sound, taste, smell, and touch.\footnote{Our sensory organs perceive signals that are proportional to the logarithm of the intensity.  This is known as the Weber-Fechner psychophysical law \cite[][p. 73]{Herrmann1984}.} The situation is analogous to Plato's allegory of the cave, where prisoners who have spent their whole lives chained in a cave see shadows cast on a blank wall in front of them.  They cannot see the actual, physical objects casting the shadows. In that sense, the observations are about the prisoners' perception of reality through their senses rather than the actual reality \citep{Denby1996}.

The question of perception versus reality is particularly challenging when dealing with observations of truly unknown objects. One feature of the human brain is that if we think we know what we are looking at in the sky with our naked eyes (a bird, balloon, or airplane of a particular linear size), we obtain a crude estimate of the object's distance. Nowadays, one might encounter a small drone 15 cm in diameter in the sky, or a much larger one. Obviously, if an observer thought he or she was seeing a four-engine jet airplane, but it was really 25 cm in length, any estimate of the object's distance would be greatly in error. It is also possible to be fooled when estimating the angular size of an object. The Sun or full Moon viewed close to the horizon often looks ``huge.'' With simple equipment (such as a cardboard device with a 6 mm hole punched in it that slides up and down on a meter stick), one can prove, based on simple geometry, that the full Moon low in the sky is, within experimental error, the same angular size as it is high in the sky on the same night \citep{Krisciunas2009}.

While some reported UAP cases are found to have earthly explanations, not every case is resolved (and some are incorrectly resolved). This can be due to a lack of information or the genuine strangeness of an incident. As of 4 January 2024, the French association GEIPAN estimated that out of 3037 reported cases, 3.3\% remain Unidentified Anomalous Phenomena, and another 32.4\% stay unidentified mainly due to a lack of accompanying data \citep{GEIPAN2024}.

In order to collect systematic, calibrated data on UAP phenomena, academic initiatives have recently emerged following the Pentagon's 2021 Office of the Director of National Intelligence (ODNI) report \citep{ODNI2021}, see e.g., VASCO's searches for extraterrestrial probes \citep{Villarroel2022a}, \textit{Galileo project} \citep{Watters2023}, \textit{Interdisciplinary Research Center for Extraterrestrial Studies}\footnote{https://www.uni-wuerzburg.de/en/news-and-events/news/detail/news/uap-neu-im-forschungskanon-1/}, \textit{UAPx} \citep{Szydagis2023} and ExoProbe \citep{ExoProbe2023}. In September 2023, NASA's UAP panel also recommended systematic, multi-sensor searches for UAP along with a centralized reporting system \citep{NASA2023}. Also non-academics have realised interesting investigations in the topic, e.g. the Nightcrawler team \citep{Tedesco2024} that have recorded unusual infrared objects flying near Robert Moses park in New York City. Among professional scientists, the usual advice is to approach the topic in an agnostic manner: first collect well-calibrated data, categorize it, and then search for UAP among the data. The NASA study team proposed to test the hypothesis that the UAP phenomenon has a phenomenology consistent with known natural or technological causes. As such, the hypothesis is only border-line falsifiable due to the extremely broad range of possible causes for any anomaly in the sky.

In contrast to this approach, the searches for extraterrestrial probes with the Vanishing and Appearing Sources during a Century of Observations (VASCO) project \citep{Villarroel2022a,Villarroel2022c} and the recently started ExoProbe project \citep{ExoProbe2023,Villarroel2024} are hypothesis-driven projects with clear motivations to search for signatures of unknown non-human spacecraft near the Earth. The VASCO project has reported groups of transients that appeared and vanished mysteriously within half an hour on pre-Sputnik photographic plates \citep{Villarroel2021,Villarroel2022c,Solano2023}. Notably, the two most interesting cases coincide timewise with the most famous UFO observation event in the last seventy years, the Washington 1952 UFO flap \citep{VillarroelDebrief}.

No matter what the origin of the UAP phenomenon is, it has an impact on society \citep{Sol2024a,McCaw2024} and on individuals who have seen UAP \citep{Sol2024b}, regardless of whether it is a real physical phenomenon or not. 

In this paper, we define and discuss what constitutes an actual observation (see Section \ref{sec:observation}) and present examples from both astronomy and the UAP world. We focus at the beginning on reports from Project Moonwatch, that was carried out at the absolute beginning of the space age when humans just started launching satellites into space. We argue against searching for UAP in the atmosphere (see Section \ref{sec:challenges}). We also present examples of UAP observations and their claimed characteristics (see Section \ref{sec:possible}). In Section \ref{sec:taxonomy}, we discuss the issues with nomenclature and the use of ``UAP'' for defining the actual research question and why UAP report centers can rarely deliver answers. We pinpoint the most interesting and relevant research questions (see Section \ref{sec:questions}), relate to the methods used for some important astronomical discoveries (see Section \ref{sec:scientificmethod}), and based on this, suggest routes to increase the robustness of the methodology in the study of UAP.
We further introduce as a thought experiment a toy model having neuro-interface extraterrestrial probes that can help design and calibrate experiments studying the UAP phenomenon (see Section \ref{sec:toymodel}). Finally, we discuss the need for ethical considerations when studying UAP for the civilian astronomer, particularly with regard to national security concerns (see Section \ref{sec:ethics}).

In this paper, we will not review the activities of any government offices (e.g., AARO), as much of their data is classified. For historical references about UAP and government involvement, we refer readers to \cite{Swords2012} and \cite{Knuth2024}.

The goal of this paper is to present solid strategies for civilian astronomers to study UAP, with a minimum number of false positives.  The aim is to understand the nature of UAP and to advance our basic understanding of the Universe rather than to interfere with any national interests. We recognize that, as civilian scientists without security clearances, our understanding may be incomplete compared to that of researchers with access to classified information who are working on the same issue within government institutions.

\section{What constitutes an observation?}\label{sec:observation}

\subsection{General comments}

%One of the first steps to understanding what we see is by %defining what constitutes an astronomical observation. 
We define an astronomical observation as a clear, measurable detection of an object or event in the sky. This observation is characterized by its location in the sky, as well as the date and time it was made.  ``Location in the sky'' could be in the horizon system of coordinates (azimuth and elevation angle) for an observer on the surface of the Earth, or it could be in the system of equatorial stellar coordinates (right ascension and declination).

When making an observation, the scientist should record all information about the equipment used, the calibration of the equipment, and the observational conditions at the time of detection. What was the Moon's phase? Was the Moon above the horizon? Were there clouds in the sky? How sharp or blurry were known stars in an image? In other words, what was the ``seeing?'' Any uncertainties in the detection of the object of attention should be noted. A significant effort must be made to ensure that the detection is not noise or an instrumental artifact (such as a ghost image arising from a bright star just outside the field of view).  We want an authentic detection of a real object or event. Hundreds of years ago an astronomical detection was annotated on parchment or paper. Today, modern CCD images automatically contain header information relating to the observation and the observing conditions.

Depending on whether one observes a clearly defined object like a star or a galaxy or one observes astronomical transients that vary in brightness or location on the sky (e.g. supernovae, novae, flaring stars, variable stars, comets, asteroids and microlensing events), the techniques of observing vary. If one wants to determine an object's shape or size, the methods vary. The observational conditions will matter. For example, due to atmospheric turbulence, stellar images are smeared out to a diameter of one or two arc seconds at a typical mountain observatory site. The smearing is usually greater at low elevation sites. Galaxies, planets, and other objects of interest are larger in angular size than one or two arc seconds. Photometry of supernovae is complicated by the fact that these objects often occur within the central regions or spiral arms of galaxies. In those cases one should obtain template images of the galaxies once the supernovae have faded away over the span of a year.  Only after the galaxy light is subtracted can one properly calibrate the brightness of the supernovae. Thus, making an observation of an astrophysical transient may involve significant follow up.

Unlike planned observations by astronomers, UAP sightings most often occur and are seen by regular citizens (i.e. untrained observers) at random times. Even so, trained scientists have reported unusual objects in the sky for hundreds of years. A remarkable compilation of UAP observations by scientists and other explorers can be found in Vallée \& Aubeck \cite{Vallee2015}. One example is when the astronomer Charles Messier observed multiple large, dark objects moving in the sky on 17 June 1777 during a lunar eclipse. There was no indication of meteor trails, and the objects exhibited ``organized'' movement. Messier wrote, ``These may have been hailstones or seeds in the air; but they were more probably small meteorites'' \citep{Ledger1879}. The observations of a supposed satellite of Venus between 1645 and 1768 by Jean Dominique Cassini, Tobias Mayer, James Short, and others is another example \citep[][]{Anonymous1884, Clerke1902}. Today we know that Venus has no satellite, which leaves open the question: what was the mystery object observed by multiple experienced observers in the seventeenth and eighteenth centuries? 

Contemporary reports of UAP are often less rigorous than published reports by scientists. Most often, all we have is a naked eye report of an event. Other times, there is some video footage or a photo. Many times, the imagery is too blurry to confidently conclude what was seen, and basic information about the observation is often missing, making it very difficult to verify the observation as authentic despite the witness' best efforts. The transient, unpredictable nature of these events in the sky, coupled with the vast number of possible known explanations, makes it a considerable challenge to devise a tentative explanation for each event, and even more so, the {\em correct} explanation for each event. As of 2024, the situation has become even more dire due to the ease of creating fake footage using deepfake technology powered by artificial intelligence.

Perhaps the biggest challenge when using reports from witnesses is the lack of sensor calibration in the human testimonies. The great challenge for scientists wanting to study UAP is to make observations of fascinating anomalies with carefully calibrated sensors under controlled conditions.

\subsection{Examples of an astronomer's observations of unusual celestial phenomena}

As mentioned above, we assert that the people most knowledgeable about the sky are advanced amateur astronomers and professional pilots.  We pose this question: just how often do experienced observers see something anomalous in the sky? The experience of author Krisciunas is given as an example.

Apart from observing in lighted, heated control rooms at mountaintop observatories (beginning in 1982), coauthor Krisciunas has observed at least 1029 nights under the stars.  This has included: 50 nights spent finding all of the 107 {\em bona fide} objects in the Messier Catalogue; 33 nights observing meteor showers under the wide open sky for 1.5 to 5 hours per night; 578 nights dedicated to making variable star estimates; 217 nights of small telescope photometry; 62 nights observing in open domes with 0.6-m telescopes, mostly those at Maunakea and Cerro Tololo; and several other observational projects.  How many observations could be categorized as ``anomalous''?  Four.  Thus, 0.4 percent of Krisciunas's ``nights under the open sky'' involved what might be characterized as a UAP. Descriptions of these events are given below.

On the evening of 14/15 November 1969 Krisciunas set out to do some backyard observing in his amateur observatory in Naperville, Illinois, starting shortly after 4:00 UT. Strangely, there seemed to be a naked eye globular cluster in the middle of the Great Square of Pegasus.  Krisciunas knew there was no such thing. Using a 6-inch f/6 Newtonian reflector and eyepieces giving 46 and 138 power magnification, the light-emitting object appeared to be milky white and circular in shape. Over the span of several minutes it was clear that the ``nebula'' was slowly growing in angular size and the surface brightness was slowly getting fainter.  In the center of the nebula was a little blinking light. No explanation came to mind. However, three months later in the February, 1970, issue of {\em Sky and Telescope} magazine the Observer's Page section provided the explanation.  What Krisciunas and many others had witnessed was the Apollo 12 mission on its way to the Moon \citep[][]{Anonymous1970}. Beginning at 04:03 UT and continuing for 11 minutes the astronauts dumped 26.6 pounds of water overboard. Within 20 minutes the cloud of molecules had disappeared. The space capsule was 110,000 miles from Earth at that time.  There were intelligent creatures piloting the spacecraft, but they were us! 

A second curious object was observed by author Krisciunas on the night of 29/30 November 1981 while doing backyard photometry in San Jose, California. At 2:13 UT the sky and the neighborhood were lit up by an object streaking through the WSW part of the sky.  It was visible for several seconds. At that time the star BS 7229 was being measured with a photoelectric photometer. That star has V = 6.39, according to the Simbad Astronomical Database \citep{SIMBAD}.  The output of the amplifier of the photometer was being recorded with a strip chart recorder.  The telescope was tracking at an elevation angle of 49.51$^{\circ}$, azimuth 309.63$^{\circ}$, hour angle west 3:38:39.  (This was determined after the fact.) From the increase of the sky brightness in the 1.38 square arc minute diaphragm of the photometer, compared to the raw V-band signal of BS 7229, we estimate that the object had an apparent magnitude of $-$10, comparable to that of the gibbous Moon.  Two other individuals corroborated the observation.  At that time Krisciunas taught part time at West Valley College in nearby Saratoga.  One of his students, Rob Hecocks, saw the object in Carmel; he noted that the object was in the north half of the sky and was traveling west.  A fellow astronomy teacher, Tom Bullock, was flying in a small plane near Interstate 5, east of Gilroy. He estimated the object's bearing to be 290 degrees. Given that the object was seen by three independent observers at the same time from three locations, we know that this was a real object.  We presume it was a bolide or piece of space debris, and that if it reached the surface of the Earth, it landed in the Pacific Ocean.

Case three. On 30/31 December 1994 Krisciunas and observing partner M. Roberts carried out photoelectric photometry using the 0.6-m telescope at the summit of Maunakea in Hawaii.  They were descending the mountain on their way to Hale Pohaku (the mid-level facility) to sleep.  It was approximately 13:30 UT (3:30 AM) when they reached the 11,000 foot level (3350 m). They looked low in the east and saw an object as red as a red stoplight.  There was no Moon and the horizon could not be seen. They stopped the vehicle and got out to have a look at the object. It appeared to vary irregularly in brightness. If it was Venus or Jupiter, why did it appear so red and unlike any star?  Krisciunas got out his copy of {\em Norton's Star Atlas} and marked an X at the appropriate place on the appropriate map (RA 15:24, DEC $-$15$^{\circ}$). When they arrived at Hale Pohaku, they went to the library and consulted the {\em Astronomical Almanac for 1994}. It was easily confirmed that the unusual object was Venus. Using accurate coordinates for Venus, it was determined later that it was at zenith angle 90$^{\circ}$ at 13:32 UT. We note that at a mountain site with a direct view of the ocean it is possible to see bright objects at zenith angles as large as 92$^{\circ}$. Given the long path length through the atmosphere of an object at very high zenith angle, the color could certainly be much redder. Within an hour of reaching Hale Pohaku, Krisciunas and Roberts went outside and saw that the object had acquired the normal light yellow color of Venus.  On this date Krisciunas wrote in his notebook, ``Venus pulsating red at rise.'' If the hypothesis of Venus had not resolved the issue, we might have wondered if the object was the long sought ``next supernova in our Galaxy.''\footnote{Our only comparable high airmass observation is a binocular-aided visual estimate ($m_v \approx 4.7$) of the brightness of Supernova 1987A from the summit of Maunakea on 8/9 March 1987 at 5:25 UT.  The zenith angle was 89.3$^{\circ}$, and the observed color of the supernova was comparable to that of Venus viewed through a similar path length of the atmosphere. Photoelectric photometry by Dopita et al. (1988) \cite{Dopita_etal1988} indicates that the photometric color of the supernova was slightly bluer than $B-V$ = 1.10 at that time, which corresponds to the color of a typical red giant star.}

In our experience the fourth case is in a category of its own. It was a weekday, a school day, in May of 1964.  The day of the month was not recorded. The site was the playground of Lester Elementary School in Downers Grove, Illinois. We note in passing that this site is roughly 30 km south of O'Hare International Airport. It was a sunny day with blue sky. At roughly 10:15 AM Central Daylight Time 10 year-old Kevin Krisciunas noticed a grey, elliptical object smaller in angular size  than the full Moon, at roughly 50$^{\circ}$ elevation angle in the east.  The object did not have a constant azimuth and elevation angle.  It was moving side to side slightly and was trending higher in the sky. Suddenly, it moved many tens of degrees toward the south in approximately one second, then got lower and lower in the southwest, moving back and forth slightly in azimuth (which could be described as  ``falling leaf motion''). It produced no smoke, no sound, and exhibited no lights. Other grade school students out at recess saw it too.  To this day, no explanation has been found for this event.

{\em However}, we note some similarity of the 1964 event to a well observed UAP that appeared shortly before sunset on 7 November 2006 at O'Hare Airport.  It was a grey elliptical shaped craft that hovered motionless over Concourse C for a few minutes just below the 1900-foot cloud deck.  It then bolted through thick clouds, leaving a hole in the cloud cover.  It was witnessed by at least a dozen United Airlines employees.  The object made no sound and exhibited no lights.\footnote{https://en.wikipedia.org/wiki/2006\_O\%27Hare\_International\_Air\-port\_UFO\_sighting and references therein.}

\subsection{UFOs found by Operation Moonwatch observers}

Citizen science programs like the Galaxy Zoo \citep{Lintott2008,Lintott2011} or the VASCO citizen science endeavor \citep{Villarroel2022c} normally engage thousands of citizens, or even hundreds of thousands. The tradition to engage regular citizens goes back decades.  For example, in 1956 a most impressive and clever citizen science program was initiated at the Smithsonian Astrophysical Observatory. It was called \textit{Operation Moonwatch} or \textit{Project Moonwatch} and was the brainchild of Fred Whipple\footnote{Fred Whipple's department boss and previous doctoral advisor was Donald Menzel, about whom we will have more to say.} \citep{McCray2008}. The project aimed to follow Sputnik-I and newly launched satellites in the sky. Moonwatch used stations spread over the planet to create an ``optical fence'' along the celestial meridian. The amateur astronomers collected data using wide-field telescopes such as the Satellite Scope marketed by Edmund Scientific Company, which gave 5.5 power magnification and a field of view of 12 degrees.\footnote{https://www.cloudynights.com/topic/540421-modernized-project-moonwatch-telescope/}.  They tested more than thirty different designs in the field in 1956.  The professionals used Baker-Nunn cameras and in this way could photograph and track satellites \citep[][pp. 76-78,100-103]{McCray2008}.    

Jacques Vall\'{e}e has kindly made available a 141 page PDF document containing many Moonwatch reports and other material \citep{ValleeDraft}.  Copies of the original reports had been sent to Prof. J. Allen Hynek. Suffice it to say that the ``Hynek File'' is a heterogeneous set of primary documents. Some are handwritten.  Some are typewritten.  Some are copies of teletype printouts.  Some are newspaper clippings. By our count the Hynek File contains accounts of 16 of the 36 UAP listed in Table 1 of Swords (2009) \cite{Swords2009}.  Below we give the reader a small set of accounts from the Hynek File.

As one can see, there were certainly interesting observations made during \textit{Project Moonwatch}. A working draft of a funding application by Jacques Vallée does a careful analysis of the UAP found by professional astronomers in the Moonwatch project \citep{ValleeDraft}. It mentions that some reports were destroyed at the stations before reaching  Moonwatch headquarters, due to the embarrassment of reporting UFOs and the stigma. This means that the UFO sample from the Moonwatch files is far from complete, and we only have a lower bound on interesting sightings from the project.

\subsubsection{West Somerville, 1959}
This observation was made on 7 January 1959, at 23:35 UT, in West Somerville, Massachusetts, and was reported by George Burns. The report reads:

{\em ``Steady white light very high altitude.  Object came from Northeast, crossed sky to the Southwest, {\rm wiggled, made a loop [our emphasis]}, and went back across the sky to the Northeast all in a total of three minutes. $-$Not aircraft.''}

\subsubsection{San Antonio, 1959}

28 May 1959, San Antonio, Texas, report written by J. H. Wooten, based on observations by R. L. Jernigan, Harold Vagtborg, Jr., and Salvador Gomez.
This is the longest (17 pages) and one of the most interesting accounts in the Hynek file.  Extant are scans of  the original handwritten correspondence, corresponding typewritten versions of those pages, and three key drawings. The first typewritten page, dated 2 June 1959 and intended for ``J. A. Hynek only,'' is entitled, ``Observation Report of Unidentified Flying Objects.'' 

{\em Six} Moonwatch scopes were aligned on the celestial meridian: one at 75$^{\circ}$ elevation angle, azimuth 0$^{\circ}$ (i.e., due north); one at 80$^{\circ}$ elevation angle, azimuth 0$^{\circ}$; one at 85$^{\circ}$ elevation angle, azimuth 0$^{\circ}$; one scope was directed at the zenith; and {\em two} scopes were aimed at elevation angle 85$^{\circ}$, azimuth 180$^{\circ}$ (i.e., due south).  A known satellite was expected to cross the meridian at 85$^{\circ}$ elevation angle, azimuth 180$^{\circ}$, at approximately 02:31 UT.  Observations commenced at 02:15 UT.

Quoting the report: {\em ``At 02:34:30 UT a very unusual phenomenon occurred... Two of our seasoned observers, who were viewing through the only two of our 'scopes which were fixed on the same point in the sky, namely 85$^{\circ}$ altitude at azimuth 180$^{\circ}$, simultaneously witnessed u.f.o's in a manner that may be a little difficult to comprehend. That is, one man observed a fleeting object that cut into the upper left limb of his circular telescope field and after describing a smooth arc or parabolic path, left the field of view at the lower left limb.  This action took place within an estimated one-tenth of a second.  The observer was concentrating his attention on the left side of the field at the time, and was certain only of what transpired on that side of the view field...[T]he other observer on the other telescope which was set to cover the identical field was concentrating his attention on the right side of the field, when he, simultaneously, observed a similar action at that side of his scope.  That is, the object entered the upper right limb, curved hyperbolically, and then left the field at the lower limb, and all within a split second of time...[T]he two objects were approximating a collision course, when, by {\rm deliberate control or mutual repulsion [our emphasis]} they veered off and went each on his merry way.''}

At the very time that the two objects were observed in the scopes aimed at 85$^{\circ}$ elevation angle, azimuth 180$^{\circ}$ the observer aimed at the zenith saw an object pass through his field on a parabolic arc. ``It occurred in the lower right quadrant.  When he sketched it on paper according to the image his power of recollection brought to mind [the report writer then] tried to match it to the configuration that was observed on the right side side of one of [the] 85$^{\circ}$ scopes.  It does not 'dove-tail' into the scheme perfectly but it probably casts some illumination on what the object was doing just before it passed into the 85$^{\circ}$ field of view.'' The information from the third observer was not revealed to the report writer until one or two days after the observations in question.

\subsubsection{Adler Planetarium, 1960}

26 to 28 August 1960, sighted in Chicago and many places in the United States.  See newspaper clipping in Hynek file entitled ``They Wonder What's Up.''  See also portion of letter of Adler Planetarium director Robert I. Johnson reproduced by  Swords (2009) \cite{Swords2009}.

Quoting Johnson, {\em ``On Friday, August 26, at approximately 9 P. M. CDT, [Adler planetarium] staff  observed a faint reddish object in the sky moving from east to west.  Apparently, this object or similar ones were almost simultaneously seen by both inexperienced and professional observers throughout the country.''} The newspaper clipping indicates that amateur astronomers from coast to coast and from Michigan to Missouri reported seeing the object.  It appeared to be about 1/10 the angular size of the Echo I balloon satellite and moving twice as fast.  Director Johnson asserted that it is {\em not} an artificial satellite or a meteor. Famous astronomer Gerard Kuiper of the University of Chicago and the University of Arizona dismissed Johnson's claims and said that the object was merely airplane lights.

\subsubsection{New Mexico, 1960}

12 October 1960, unspecified location in New Mexico, observed by Charles Capen.  Teletype printout reads: {\em ``Charles Capen observed unidentified lights travelling from north to south southwest at 0035 UT 12 October. Consisted of eight yellowish lights arranged in two triangles of three each and followed by lights in echelon travelling slow when first seen in north.  Approximate angular velocity when passing across observer's line of sight on order of 2000 sec[onds] of arc per sec[ond].  Range uncertain.''} Two thousand arc seconds per second is 0.56 degrees per second.  Charles Capen (1926-1986) was an experienced observational astronomer who was the Association of Lunar and Planetary Observers' Mars Section Recorder for 17 years.\footnote{chrome-extension://efaidnbmnnnibpcajpcglclefindmkaj/https://alpo-astronomy.org/jbeish/ChickCapen.pdf}  Telephone contact with Capen revealed that he believed the objects were high flying aircraft.  This agrees with analyis by ATIC (Air Technical Intelligence Center).
    
\subsubsection{Bedford, 1961}

14 June 1961, Bedford, Massachusetts, observed by James L. Vanderveen, M. D.  Time - between 3:00 and 3:05 AM. 

\textit{``Direction Low in northeast sky, passing from north toward east; at maximum height in its arc it appeared to be about
the [height] of two fingers held out at arm's length,
above a high horizon (tall trees 200-300 feet distant).''}
[Thus, the object was about 15$^{\circ}$ above the tree line.]

{\em ``Intensity.  Roughly as bright (at its most intense)
as a full Moon which has risen 1/4 of the distance
between horizon and zenith under conditions of good
visibility.  Color was roughly of the same quality. [Unclear what that means.]  The brilliance was variable and appeared to have a regular period of ... 1 second.  Light was still visible at the low point of intensity.''}

   {\em ``Length of observation: about 30 seconds, during which
time it traversed the aperture of a 30 inch wide window
placed 8 feet from observer's eye.''}  [The window subtended
an angle of about 18 degrees.]

The apparent diameter of the object was about 1 cm at arm's length.
For a 62 cm arm length that corresponds to an angular diameter
of about 0.9 degrees, which is roughly twice the
angular diameter of the full Moon.

\subsubsection{Eastern Airlines, 1966}
 
 18 May 1966, 09:40 [P.M.] Eastern Standard Time, over the Gulf of Mexico at latitude 24$^{\circ} 00'$ N, longitude 96$^{\circ} 00'$ W, reported by the S/O [safety officer], D. Leppard.  

The safety officer for EAL [Eastern Airlines] flight 905 reports: {\em ``[An] object appeared at first as a short, vertical blue or blue-green line.  It moved from right to left becoming circular until it reached approximately the size and brilliance of the Moon. After approximately one minute it disappeared, leaving a bluish glow about five times the size of the Moon for an additional 3 minutes.''}

Information is handwritten on a form entitled, ``Flight Crew Member Report Form for Satellite Re-Entry or Bright Fireball in conjunction with Smithsonian Astrophysical Observatory's Re-Entry and Recovery Program.''  ``U.F.O.'' is written in red pencil at the top of the page.  It is not specified if the time is A. M. or P. M., but from the context it must have been night.  There was no Moon visible because new Moon occurred within a day of the sighting.

The flight was en route to Mexico City.  The plane's magnetic heading was 225$^{\circ}$, true heading 233$^{\circ}$, and the altitude was 31,000 feet.  The object appeared 25$^{\circ}$ to the right of the heading of the plane, and 20$^{\circ}$ above the horizon.  The names of the captain and the ``F/O'' [first offficer] are given, but there is a note: {\em ``Captain and F/O do not wish to be identified with sighting.''}

\section{Arguments against searching for UAP in the atmosphere}\label{sec:challenges}

Consider that the whole sky contains $4 \pi (180/\pi)^2 \approx 41253$ square degrees. To search for \textit{any} fast-moving objects we need to monitor large swaths of the sky. One interesting way to experience this vastness is to lay out on a chaise lounge after midnight on a (preferably moonless) night when a meteor shower is active (such as August 12, November 17, or December 13 most every year). This requires no telescope. In fact, a telescope would be counterproductive. One cannot know if the next meteor will appear tens of degrees to the right, left, above or below the meteor shower radiant. 

Historically, one used photographic plates (like those used for the National Geographic Society, Palomar Observatory Sky Survey of the 1950s for mapping the sky. In order to make an atlas covering the entire sky using 6 by 6 degree photographic plates would require a minimum of 1146 plates. There should be a certain overlap of the fields, so perhaps the number would increase to 1500. Information on the colors and temperatures of stars can be obtained by taking two sets of plates, one using a blue filter and one set using a red filter. Finally, because some constellations are always below the horizon depending on one's latitude, an all-sky survey would need one telescope in the northern hemisphere and another one in southern hemisphere.

Modern astronomy often uses CCD (charge coupled device) devices instead of photographic plates. CCD imagery at an astronomical observatory generates images containing header information such as date, name of target, start time of observation, length of exposure in seconds, name of telescope, latitude and longitude, filter name, elevation of object above horizon, azimuth of object, hour angle of object, air mass value of observation, and plate scale of camera-plus-telescope combination (measured in arc seconds per pixel), which is determined from fields of stars with accurately known stellar coordinates.

Many CCD cameras on 1-m class telescopes used by author Krisciunas since the mid-1990s gave a field of view on the order of 8 by 8 arc minutes. One arc minute equals 1/60 of a degree. So the field of view of such a camera is only 0.018 square degrees. One would never attempt to produce an all-sky survey using such a camera on a 1-m telescope, as it would take many lifetimes.

Another thing to consider is the dynamic range of a telescope, namely, how bright is the brightest object one can image with a short exposure while not saturating the detector, along with the limiting magnitude for a 5-$\sigma$ detection above the night sky level of a point source. Telescopes are designed to detect the faintest possible objects because at least when it comes to extragalactic sources this maximizes the volume of space one can study, which in turn maximizes the number of potential objects to study. Using CCD images on a 1-m telescope and short (3 second) exposures we can do photometry on ``bright objects'' (apparent magnitude 10). Brighter objects such as UAP visible to the naked eye (brighter than magnitude 6) would saturate the CCD.  The faintest stars that can be accurately measured with such a telescope are about apparent magnitude 20.

The Vera Rubin Observatory, now nearing completion at Cerro Pachon in Chile, has an 8.4-m diameter primary mirror, with a light gathering power equivalent to a 6.4-m diameter unobstructed mirror. Its 3200 megapixel camera covers 9.6 square degrees on the sky and will be used to image the entire sky visible at that latitude every 3 to 4 nights, producing massive amounts of data on stationary stars and galaxies, moving objects (such as asteroids), and objects of variable brightness (such as supernovae). In single images the faintest stars detected will have an $r$-band magnitude of 24.5.\footnote{https://en.wikipedia.org/wiki/Vera\_C.\_Rubin\_Observatory} The brightest non-saturated stars will be 17th magnitude.

An obvious solution for the detection of bright UAP is to use small cameras instead of telescopes, allowing all-sky coverage in one single exposure. However, this is a needle-in-the-haystack approach when applied to the search for ``Unidentified Anomalous Phenomena.'' The clearest challenge to such an exercise with focus on the ``unidentified'' comes down to the number of false positives following this needle-in-the-haystack approach. Unfortunately, this haystack, the Earth's atmosphere, contains literally hundreds of thousands of needles that are not UAP. This haystack contains hypersonic aircraft, surveillance aircraft from every wealthy country, many thousands of military drones, military and weather balloons of all sizes, colors, and trajectories. Finally, one must consider occasional missiles 
both within the atmosphere and sometimes ballistic ones, re-entering the haystack from outer space anywhere in the world.

These many thousands of decoy ``needles'' are being invented anew each year, in secret, by military forces that purposely hide them from all adversaries. Indeed, we do not even know the new, classified, military aircraft invented by our own countries. The newly invented aircraft have different sizes, shapes, colors, altitudes, and speeds compared to ``needles'' previously created by humanity. Any artificial intelligence system trained on a recent set of ``needles'' in the atmosphere will be fooled by new ones.

The defense forces of a country, on the other hand, must develop hardware and software to detect and identify fast moving objects in the sky, as they could be missiles from hostile actors. Intercepting them is a matter of life and death. Billions of dollars are spent on such systems each year. It is these kinds of systems that one might want for the detection of UAP, but professional and amateur astronomers cannot compete with military forces in this regard. Because information relating to military systems would be classified, here we make no attempts to evaluate the capabilities of such systems.

Therefore, the simplest solution to this dilemma for the civilian astronomer is to look for transient sources \textit{outside the atmosphere}. This is the domain where astronomers have unique experience regarding the detection of interesting objects, characterizing them, and measuring their distances. This will also spare the astronomer procedural conflicts due to ethical concerns when entering the domain of military interest regarding searches for UAP.

Given how difficult it is to observe fast-moving objects with a professional telescope, we can now understand some of the basic reasons why present-day automated surveys like the Zwicky Transient Facility and the Sloan Digital Sky Survey are unlikely to capture a UAP in any convincing manner. Small mobile cameras, military sensors, and the naked eye are better equipped to study UAP in the atmosphere, as are meteor cameras. This stands in clear contradiction to the common claims by some professional astronomers that their favorite transient surveys, which detect thousands of transients every night, would have confirmed the existence of flying saucers long ago if they were real.

\section{A possible physical phenomenon}\label{sec:possible}

As we mentioned in the introduction and flesh out in Section \ref{sec:taxonomy}, the UAP definition is an over-broad category, a ``junk drawer'', and in many ways refers to Temporarily Non-Attributed Objects (TNOs; see Karl Nell's talk at the Sol Foundation) which are false positives. But there are certain properties commonly assigned to UAP, that can help to distinguish them from TNOs. We note that there is a lack of openly accessible, published scientific evidence supporting any of these properties. However, these properties were observed during events such as the \textit{Nimitz} encounter in 2004, which remains one of the most elusive encounters to date. In this incident, multiple military personnel reported visual sightings of vehicles with extraordinary capabilities and recorded them on their instruments\footnote{Presumably, the U.S. government has classified much of the material from the \textit{Nimitz} incident, which could provide strong evidence.} \citep{Knuth2019}. UAP manifest Five Observables \citep{Elizondo2024}:

\begin{enumerate}
\item \textbf{Sudden and instantaneous accelerations.} Such accelerations sometimes display maneuvers that no airplanes, drones, rockets, or flying animals are capable of. If real, they would defy any technology created by humans.
 
    \item \textbf{They fly although they have no obvious flying capabilities.} Many times, the objects show no signs of propulsion, nor do they have any wings, rudders, or other features required by aerodynamics -- yet they still fly! Not only that, but sometimes these objects swiftly change trajectories.

    \item \textbf{No sonic booms, fireballs, etc.} This is self-explanatory: the objects can move at extraordinary velocities without the usual associated signatures. For example, an object entering the atmosphere without producing a fireball is, of course, greatly interesting.
  
    \item \textbf{Transmedium capabilities.} Here, we discuss artificial objects that can pass from vacuum to air and then to water.
 
    \item \textbf{Low observability.} This refers to the common ``blurry dot'' appearance of UAP. Why is it so difficult to make a sharp image of a UAP?
\end{enumerate}

Some of these properties might be misidentified optical illusions. Confusing factors related to parallax, perspective, image resolution, or video quality can further create the impression of a UAP \citep{AARO2024B}. Other properties might result from a selection effect; for example, the ``blurry dot'' problem alone warrants an alternative explanation. If the nomenclature favors 'flying saucers' over UAP, as it did in the 1950s, people may be less likely to report anything that does not resemble a flying saucer -- the broadness of the UAP category might contribute to the prevalence of blurry dots. Additionally, sharp images may be perceived as fake. Until a proper scientific investigation is conducted in search of non-human aircraft, we cannot know the true extent of the ``blurry dot'' problem and whether it indicates an exaggerated phenomenon or a relevant observational property.

\subsection{The Washington D.C. 1952 Flap}
The UFO literature has documented numerous significant incidents where the five properties listed above have been present in various forms \citep[e.g.][]{ValleeFS1,ValleeFS2}, some of which are difficult to dismiss as misidentifications.

A compelling example is the famous \textit{Washington D.C. 1952 UFO flyover} (also known as the \textit{flap} or \textit{carousel}). To date, this remains the most well known mass sighting, resulting in thousands of news items and the largest press conference since the end of World War II, held by the U.S. Air Force on 29 July 1952. During two consecutive weekends, 19-20 July and 26-27 July 1952, numerous UFO sightings were reported. Pilots, stewardesses, and airport radar controllers at Washington airports simultaneously observed multiple UFOs, some of which exhibited unusual movements and accelerations. Fighter jets were dispatched to intercept the UFOs, and one jet even reportedly shot down a UFO \citep{NICAP2007}. Reports indicate that about 35 times as many UFOs were reported in July 1952 as at any other time \citep{Australia}.

The U.S. Air Force dismissed the event as a series of misidentifications made by experienced pilots and radar controllers. Assisting them was theoretical astronomer Donald Menzel, who provided the U.S. Air Force with the weather disturbance idea to explain the radar observations. However, this explanation was later shown to be incorrect by meteorologist James E. McDonald.\footnote{James E. McDonald and Donald Menzel became sworn enemies, leading to many intense attacks on McDonald's persona. Tragically, McDonald died by suicide in his early fifties.}

Independent of the intriguing stories surrounding the event, there may be an additional, independent source of support for the Washington 1952 UFO observations. Interestingly, the two most improbable and brightest VASCO multiple transient candidates reported—a triple transient \citep{Solano2023} and five aligned transients \citep{Villarroel2022c}—coincidentally occurred on 19 and 27 July 1952, respectively.\footnote{The original date of 28 July 1952 in Villarroel et al. (2022c) \cite{Villarroel2022c} was incorrectly tabulated; it should be 27 July, as clarified by Villarroel (2024, Debrief) \cite{VillarroelDebrief}.} These two examples were found and documented before the authors were aware of the flap.

The two independent coincidences become even more intriguing when considering that in the autumn of 1952, the newly appointed Harvard Observatory director, Donald Menzel -- the same person who helped the U.S. Air Force explain the Washington 1952 UFO flap -- decided, as one of his first actions in the role, to destroy roughly one-third of the Harvard Observatory plate archive, the largest photographic plate collection in the world. This episode is carefully documented in the American astronomer Dorrit Hoffleit's memoir, ``Misfortunes as Blessings in Disguise'' \citep{Hoffleit2002}. Hoffleit describes how Menzel negatively impacted her career after she attempted to save the plates. Between 1953 and 1967, Menzel also halted the Harvard astronomers' work on the astronomical survey, an episode referred to by Harvard as the 'Menzel gap' \footnote{https://hco.cfa.harvard.edu/about/}. It is, of course, interesting to consider whether there might have been some support for the Washington D.C. 1952 flap in the plates that the U.S. Air Force ordered to be destroyed (or misplaced). Notably, Donald Menzel had the highest security clearances within the National Security Agency.\footnote{Menzel's NSA connection was revealed by Stanton Friedman after his death. Menzel's name is also associated with the most famous UFO ``conspiracy theory'' of all time—the so-called Majestic-12 documents, which describe a committee allegedly responsible for a UFO cover-up, of which Menzel was claimed to be a part. The MJ-12 documents come in several collections and are highly controversial, with the U.S. Air Force asserting that they are false. Menzel was a notorious UFO skeptic and even became a part of the most famous UFO conspiracy theories.}

\section{Lost in the taxonomy forest}\label{sec:taxonomy}

\textit{``What's a Good Name? [...] When a field naturalist made a discovery, he first identified the find as something new or a variant of organized knowledge. He then classified it, and gave it a descriptive name. Now when we discover an unconventional object, we identify it as “unidentified” and name it the same! On the first page of The Report on Unidentified Flying Objects, we find whom to thank for this contradiction. Major Edward Ruppelt says, “UFO is the official term that I created to replace the words flying saucers.” One suspects that a field naturalist would have done considerably better, as naming was their specialty. Ruppelt scored a complete miss on two out of three words: unidentified and flying. It is assumed that anyone with a good dictionary can see why unidentified is a misnomer.''}

This quote is from former NASA employee Paul R. Hill's book \citep{Hill2015} and refers to the unreasonable nomenclature used to define the phenomenon. Both 'UFOs' and 'UAP' are dumping or leftover categories of objects that fit no known class of objects, but they are not any new class of objects by their own merit. Unlike the category 'flying saucers' that have distinct features and can give rise to concrete predictions for what to look for in surveys, 'UAP' do not provide any predictions and cannot be used for hypothesis-driven science (see Section \ref{sec:scientificmethod}).

One may compare this proposed agnostic method to  Hubble's scheme for classifyinging galaxies based on morphology and colors \citep{Hubble1936}. Similarly, Carl von Linné (Linnaeus) invented modern scientific taxonomy by categorizing living beings (animals, plants) \citep{Linnaeus}. The argument for UAP taxonomy is that one may learn about the most common types of misidentifications and the most interesting classes of remaining unknown objects that cannot be easily identified. The erroneous comparison between categorizing UAP and categorizing galaxies is that when Hubble or Linnaeus categorized their objects, their goal was not to understand the ``unidentified'' but to provide basic classification schemes of all known phenomena. These were entirely different goals.

Nevertheless, UFO report centers worldwide have done historically valuable work in collecting and categorizing UAP reports. These groups include MUFON, Archives for the Unknown\footnote{https://www.afu.se}, French Groupe d'Études et d'Informations sur les Phénomènes Aérospatiaux Non-identifiés (GEIPAN), and many other local organizations all over the planet. Each report is carefully studied with help of field investigators that search natural explanations to each case. Unexplained cases are classified as a UAP. These reports are not without value, even if the data collected often is heterogenous and lacking in standardization (humans are uncalibrated sets of sensors). For particularly good cases, where witnesses and imagery complement each other, one can do careful case studies. Examples of UFO report centra that apply the scientific method to study particularly interesting UFO cases are the Scientific Coalition of UAP Studies\footnote{https://www.explorescu.org} and the 3AF Sigma-2 society\footnote{https://www.google.com/search?client=safari\&rls=en\&q=3AF+Sigma-2\&ie=UTF-8\&oe=UTF-8} , where physicists and scientists from other fields apply numerical simulations and methods of physics to test various possible explanations for a UAP case. 

In UAP research, it is not uncommon for serious amateurs and professional scientists alike to discuss the need for ``agnostic'' research on UAP. According to this approach, scientists are advised to ignore ``unreasonable ideas'' such as the extraterrestrial hypothesis due to the difficulties with interstellar travel \citep{Frank2024}. Instead, they often propose working with untargeted data collection and categorization using very careful, systematic methods with the goal of identifying unknown objects in the sample. This is done with a good, predefined knowledge of the instruments used and with an established robust data collection setup to minimize biases and background contamination, and maximize completeness. Instruments are well calibrated, allowing homogenous data collection. After the data collection stage has been completed, the data are carefully analyzed, all known objects identified (possibly with artificial intelligence or other automated methods), employing the exclusion method. Finally, any outliers are identified.

In Section \ref{sec:scientificmethod}, we propose more convenient methods of data collection.

\section{What are the most urgent questions in UAP research?}\label{sec:questions}

The question, ``What is the origin of UAP?'' is not conducive to a single hypothesis-driven experiment. However, several important questions can be investigated individually. These questions relate both to the origin and the potential effects of UAP. Here, we outline some key questions worth exploring:

\begin{itemize}
  \item ``Do UAP pose an aerial or maritime hazard?''. This question can be addressed by examining whether there is a higher incidence of airplane or boat accidents near recognized UFO hotspots. While this investigation can enhance our understanding of the phenomenon's impact, it will not provide direct insights into its origin, whether natural, human-made, or non-human artificial. This question is of interest to both civilian and military scientists.
  
  \item ``Do UAP pose a health threat?'' This can be studied through comprehensive health examinations of individuals with recent exposure to UAP phenomena. Medical coalitions such as uNHIdden, a London-based mental health and wellbeing organisation, could support this research.
  
  \item ``Are there non-human artifacts on the Earth? What are typical properties of such artifacts?'' This effort is complex and would either require the declassification of materials of undetermined composition from companies such as Lockheed Martin\footnote{https://deadline.com/2021/04/ufo-fragments-in-possession-lockheed-martin-says-harry-reid-1234748095/} and Northrup Grumman\footnote{https://thehill.com/homenews/senate/4097653-senators-to-offer-amendment-to-require-government-to-make-ufo-records-public/} \citep{Elizondo2024}, or a civilian-led crash retrieval initiative similar to the European Crash Retrieval Initiative.\footnote{www.ecr-initiative.org} A discovery would provide an immediate and definitive answer to whether the claims of crashed flying saucers are true. Some interesting anomalies worth further study include the Baltic Sea Anomaly\footnote{https://science.howstuffworks.com/environmental/earth/oceanography/baltic-sea-anomaly.htm} discovered by Ocean X's team in Sweden \citep{OceanX}, an anomaly at the bottom of the ocean outside the Swedish coast with an unknown origin. The substructures of the big round object look like a staircase. There are straight lines and right angles. Various hypotheses have arisen, from crashed UFOs to structures built by earlier civilizations. This anomaly lost the public's interest when geologists reported the material to be made out of ordinary granite. Comments by the geologists were made directly to the media\footnote{https://www.livescience.com/22846-mysterious-baltic-sea-object-is-a-glacial-deposit.html} without any prior scientific report or published analysis. The unfortunate detail was that the divers did not bring samples from the anomaly itself as the object was too hard to cut samples from. Instead, the divers brought samples from the surroundings, which were made of granite.\footnote{Carl Sagan's famous sentence, ``Extraordinary claims require extraordinary evidence,'' often means that anyone can criticize the extraordinary claim without holding their own analysis to any standards of quality.} As such, the object still merits further analysis as its material composition remains undetermined.
  
  \item ``Are there non-human artifacts near the Earth?'' Since 2021 searches for non-human artifacts have been conducted by the \textit{Galileo Project} and the VASCO Project. This hypothesis is concrete, allowing for testable predictions.
  \item ``Can we predict when and where is the next UAP observation?'' Understanding why and where these phenomena appear would greatly enhance our understanding of their nature.
  \item ``Are some UAP a kind of earthquake light?'' Recent studies suggest that some phenomena, such as the Hessdalen lights \citep{Hessdalen}, may be related to earthquake lights. Or could they be a form of ball lightning or weather-related phenomena?

\end{itemize}

While some of these questions can enhance our understanding of the UAP effects, they may not all contribute to fundamental scientific progress. For astronomers it is essential to review the most relevant problems related to UAP research. 

For example, consider the question of non-human objects. Can we find signs of non-human objects outside Earth's atmosphere but inside the Solar System? This could include artificial objects that for example have movements or orbits not matching any known human objects, but also other ``entities'' with unusual spectral energy distributions. Can we detect objects entering Earth's atmosphere without experiencing combustion?

Further, do some unusual fast transients have a temporal connection with known UAP sightings? Previous work \citep{VillarroelDebrief} has suggested that some extraordinary cases of multiple transients were detected during the Washington 1952 flap. As authentic multiple transients are best explained by an artificial origin, such a correlation is of serious interest.

There are even more speculative theories to examine what we have already proposed. Do we observe any signatures of warp drives or wormholes in astronomical surveys? One approach is to examine the sky in the region of modern UAP sightings with good constraints in location and examine the objects' positions in search for small distortions in the object coordinates. This could be facilitated by surveys such as the Gaia survey, which in principle could detect coordinate shifts (fractions of arc seconds) due to gravitational waves. Naturally, one would have to examine all potential changes in coordinates within a box of many degrees where the UAP is believed to have been seen, as the UAP location is typically poorly constrained. However, the exact time of the sighting might be accurately known.

\section{Setting the scientific method to practice}\label{sec:scientificmethod}

Since the 17th century, the scientific method has continuously evolved to develop the most efficient means of progressing science. Through the hypothetico-deductive method, which closely aligns with the ideal scientific method, we recognize the importance of formulating a \textit{falsifiable hypothesis} \citep{Popper1959}. 

The scientific method requires scientists to identify the most important questions to investigate, often based on a model or theory. This model, theory, or plain hypothesis allows us to make predictions. Once a prediction is made, an experiment to collect empirical or simulated data is conducted. The data are then analyzed, and a conclusion is drawn regarding whether the hypothesis has been falsified. If the predictions are not fulfilled, the null hypothesis is considered \textit{falsified} \citep{Kuhn1962}. If the results are difficult to interpret, the hypothesis may need to be refined, and the predictions sharpened. In practice, additional {\em ad hoc} assumptions are sometimes introduced to account for various scenarios where the predictions may fail even if the model, theory, or hypothesis is correct, leading to repetition of the experiment.
However, if the predictions based on the hypothesis are validated during the test, this provides additional support for the model or theory. 

The scientific method stands in direct contrast to the agnostic approach, as proposed by many skeptics (see Section \ref{sec:taxonomy}). A common misconception is that a plausible hypothesis must also conform to \textit{common sense}. While common sense can guide the selection of interesting hypotheses, it is also susceptible to prejudices, biases, and dogmatic thinking when evaluating data. In astronomy, observations lead to the formulation of hypotheses and theories, and new observations help to falsify, modify, or strengthen existing models.

\subsection{Historical examples in astronomy}\label{sec:history}

A good example of hypothesis-driven science involves the expanding universe. Between 1929 and 1998, astronomers had adopted the hypothesis that the rate of expansion of the universe must be decelerating due to its gravitational mass. Galaxies are gravitationally attracted to each other, causing the velocities between them to decelerate. The dominant research effort in cosmology was determining the rate of expansion, known as the Hubble constant, and the rate of deceleration of the expansion of the Universe. 

In more detail, the hypothesis involved two vital properties of the universe, the rate of its expansion, $H_{0}$, measured in km s$^{-1}$ Mpc$^{-1}$,
and the average mass density of the universe. Those two properties led to the hypothesis that the universe must be decelerating in its expansion owing to the mutual gravitational attraction of all the matter in the universe, which we now assume are of two different kinds, baryonic matter and the Dark Matter, where the existence of Dark Matter still has to be demonstrated\footnote{According to the Modified Newtonian Dynamics (MOND) theory, one could argue that dark matter may not be necessary. So far, CERN has not confirmed a single candidate particle for dark matter.}. The hypothesis of deceleration was rooted in our understanding of Newtonian gravity and General Relativity.  It was also intuitively logical. 

The hypothesis of deceleration led to three possible fates for the universe. If our universe has an overall mass density above some critical value, the galaxies would eventually gravitationally ``fall'' back toward each other. If our universe has a density below that critical value, the universe will expand forever. The hypothesis includes a third possible fate, that the universe will ultimately coast to a stop and will neither expand nor contract. The hypothesis of a decelerating universe spawned these profound questions about the fate of the universe, and it dominated cosmology for seven decades. For more details, introductions to cosmology during that era are provided by Kolb \& Turner (1990) \cite{Kolb1990}, Carroll, Press \& Turner (1992) \cite{Carroll1992}, and Ryden (2003) \citep{Ryden2003}.

To test the hypothesis, challenging searches for pulsating Cepheid variable stars located in the Virgo and Coma galaxy clusters were carried out with the world’s largest telescopes, including the Hubble Space Telescope \citep{Freedman_etal2001}, to determine accurately the rate of expansion, the Hubble constant. They found the Hubble constant to be 72 $\pm$ 8 km s$^{-1}$ Mpc$^{-1}$.

Meanwhile, the hypothesis of deceleration and the rate of deceleration remained untested. Two teams of observational astronomers embarked on major observational programs to use Type Ia supernova explosions as ``standard candles'' to measure the deceleration of the expansion of the universe. In 1998, both teams found their measurements showed the hypothesis of deceleration was wrong \citep{Riess1998,Perlmutter1999}. Their measurements showed that the expansion of the universe was accelerating, not decelerating. The specific and long-lived hypothesis of a decelerating universe motivated incredibly complex and difficult measurements of that deceleration, ultimately showing that the hypothesis was not only incorrect but that another phenomenon, Dark Energy, now dominates the dynamics of the universe. Without the hypothesis of deceleration, spawning painstaking measurements with the world’s largest telescopes, the discovery of Dark Energy might have happened very differently.

A second example comes from the world of exoplanets. Since the time of the ancient Greeks, philosophers have hypothesized about the possible existence of planets orbiting other stars and the possible life on them. Epicurus (341-270 BCE) wrote, ``There is an infinite number of worlds, some like this world, others unlike it.'' In 1734, Emanuel Swedenborg \citep{Baker1983}, a Swedish theologian and scientist, was the first to hypothesize that our Solar System, with its Sun and six known planets, formed from a rotating cloud of gas and dust that flattened and collapsed by gravity. Kant and Laplace embellished that ``nebular theory'' of the formation of planets around the Sun and other stars (e.g. \citep{Woolfson1993}). This planet formation hypothesis received support from the discovery of young Sun-like stars in the 1940s, the ``T-Tauri stars'', around which platter-shaped ``disks'' of gas and dust were found (e.g. \citep{Hartmann1998}). Still, the hypothesis of planet formation and the physics-based models could not predict how often planets emerged. Were they rare ``one in a thousand'' outcomes or common? The hypothesis and models also could not predict the diversity of planet masses, orbits, and chemical compositions, never mind water-bearing habitability. 

The planet formation hypothesis motivated several groups to develop a new technique to detect planets around other stars. This took a decade. Their idea was the same: to measure the Doppler shift of the light from a star ultra-precisely.  Since a star and a planet both orbit the center of mass of the star-planet system, one might be able to deduce the existence of such a planet from small variations in the radial velocity of the star.  In 1995, the first three planets were discovered \citep{Mayor1995,Marcy1996}. Within four years, 28 ``exoplanets'' had been discovered \citep{Marcy2000}. The hypothesis of planet formation around young stars was vital in motivating several teams to dedicate their careers to developing a specialized technique that enabled the discoveries. Just pointing powerful telescopes at other stars would not have succeeded.

In a similar fashion, other important discoveries in modern astronomy have been made. Hypotheses concerning the formation of the Milky Way were formulated after the discovery of quasars and after the understanding that quasars, extremely luminous engines found in some galaxy cores, are powered by the accretion of hot gas by supermassive black holes. After Balick \& Brown (1974) \cite{Balick1974}
observed strong radio emission from a compact source in the center of the Milky Way, namely Sagittarius A* (Sgr A*), it led to the suspicion that the region hosts a supermassive black hole. Two competing teams, one led by Andrea Ghez at University of California, Los Angeles, and the other by Reinhard Genzel at the Max Planck Institute for Extraterrestrial Physics in Germany, set out to track the positions of stars in the Galactic Centre region by means of high-resolution near-infrared imaging.  The motions of stars near Sgr A$^*$ were tracked over several years. Stars were observed to move in orbits around an invisible point of mass $\sim$ 4 $\times 10^{6}$ $M_\odot $. In this way, the two teams discovered a supermassive black hole in the center of the Milky Way. See Ghez et al. (2005) \citep{Ghez_etal2005} and Genzel, Eisenhauer \& Gillesen (2010) \citep{Genzel_etal2010}. 

Serendipitous discoveries have in many cases been among the most fascinating ones in science. When Arno Penzias and Robert Wilson at Bell Labs discovered a persistent microwave signal that was detectable no matter which direction they pointed their radio antenna, they serendipitously discovered the Cosmic Microwave Background Radiation -- a residual afterglow of the Big Bang {\citep{Penzias_Wilson1965}. Uranus {\cite[][pp. 42-47]{Sheehan2021}, pulsars \citep{Hewish_etal1968}, quasars \citep[][]{Kellermann_Bouton2023},
and Fast Radio Bursts \citep{Lorimer_etal2007} are all examples of serendipitous discoveries in astronomy. Serendipitous discoveries by definition cannot be planned for.  Yet, as Louis Pasteur said in a notable speech,\footnote{Pasteur was undoubtedly speaking in French.  An equivalent translation is,  ``\textit{Chance favors the prepared mind.}'' See: https://www.pasteurbrewing.com/louis-pasteur-chance-favors-the-prepared-mind/} ``\textit{Fortune favors the prepared mind.}'' It follows that hypothesis-driven science is the most efficient way of progressing in new fields.

%Discuss how the greatest discoveries in modern astronomy were built on hypothesis-driven science. Examples with exoplanets, dark energy,... 

\subsection{Experimental design}\label{ref:design}

Experimental design should ideally reflect the research question at hand. To 
illustrate this, consider the example of the discovery of Dark Energy discussed 
in Section \ref{sec:scientificmethod}. Specifically, would the acceleration of the Universe's expansion have been discovered if data were collected from a supernova report center based on testimony and pictures taken by amateur astronomers, similar to the heterogeneous data collection methods used by UAP report centers?\footnote{For argument's sake, let us assume that amateurs had access to 4-m class telescopes. This was necessary to reach a redshift of $z \sim 0.5$.} 

Such a collection of supernova reports would suffer from variations in data 
quality and observational conditions, and there would be no way to control 
selection effects, making it very difficult to draw reliable conclusions due to 
uncontrolled uncertainties and biases. Instead of accurately measuring the 
deceleration parameter over a wide range of redshifts, this approach would result 
in a heterogeneous sample with various redshift bins and biases, possibly including 
different types of supernovae, not just Type Ia SNe, thereby complicating the 
analysis. It is unlikely that the acceleration of the universe's expansion 
would have been discovered under such conditions. Similarly, it is unlikely that 
any serious hypothesis concerning the nature of many UAP can be studied through 
largely heterogeneous samples, such as those derived from UAP report centers. 
Systematic answers require systematic approaches with very clearly defined goals.

Another principle to adhere to is \textit{reproducibility}. It is possible to 
argue that a particular UAP might be seen, but never seen again. The same is true for any supernova that explodes.  The event will never happen a second time. We might not be able to observe the behavior of a particular anomalous object several times, but we might be able to find several similar objects during a systematic search carried out over years. This is exactly what was done by \cite{Krisciunas_etal2000}, who used a small amount of optical and near-infrared photometry of several Type Ia supernovae to construct light curve and color curve templates, allowing one to fill in gaps in the record for more sparsely sampled objects. The goal was to obtain optical and near-infrared photometric maxima of as many Type Ia supernovae as possible.\footnote{The combination of optical and near-IR data leads to more robust corrections for any dimming and reddening effects of interstellar dust along the line of sight. This decreases the systematic errors of the distances to invidivual supernovae and their host galaxies.}  Three years later, \cite{Krisciunas_etal2003} vindicated their previous prognostications based on better sampled supernova light curves and physics-based supernova modeling.

Detecting duplicate or multiple examples of a particular object with similar properties is, indeed, as good as reproducibility. In this way we can fulfill the expectation of \textit{reproducibility} of a sighting so that it can be understood, having reasonably full control over the conditions under which the anomaly was sighted. This is much more difficult to fulfill if only relying on UAP reports which, at first, might be tainted by reporting bias.

Any detection will need to have a \textit{good signal-to-noise ratio} and any UAP search project will need to use instruments with good dynamic range. This is because nighttime telescopes are generally built to look for faint objects, i.e. objects near the detection limit. Very bright objects like airplanes or a very bright UAP inside the atmosphere will easily saturate the image. One single instrument is unlikely to be good enough in most cases.

In their 2023 report, the NASA UAP panel \citep{NASA2023} strongly emphasized the need to collect data using multiple instruments to assess the UAP phenomenon. For example, 
while many scientists instinctively dismiss the US Navy videos demonstrating 
nothing more than blurry dots in the videos of UAP taken by pilots, few realize 
that the blurry dots were detected by multiple sensors \citep{ODNI2021}. This means the objects detected were real physical objects. Similarly, validation with multiple, independent instruments allows us to separate artifacts (e.g. optical artifacts, idiosynchronies of the camera, cosmic ray hits) and real, 
physical objects. Both the Galileo project and ExoProbe use techniques of 
validation by using several telescopes and sensors, see e.g. \citep{Watters2023,ExoProbe2023}. 

While the data collection process helps with validation and reproduction of any finding, it does not come without challenges. The more data are collected, the more false positives one must sift through. Automated methods of removing false positives becomes crucial in such a process.

One possibility is to use artificial intelligence and machine learning to 
identify interesting outliers. It is an approach 
currently popular in the world of astronomy to search for interesting objects in vast catalogues of data. Also, several of the UAP projects use a similar approach. 

An alternative but more powerful approach is to minimize the number of false 
positives. Hypothesis-driven science methods, where a specific hypothesis is 
tested, allow the scientists to target the data collection. This means that false 
positives can be filtered out at the earliest possible stage. As an example, when 
VASCO carried out a search for signatures of extraterrestrial probes, they used 
photographic plate material \citep{Villarroel2021,Villarroel2022a} from the early 1950s, several years before 
the first man-made satellites were launched into orbit, allowing the removal of all man-made objects in space while searching for signatures of artificial objects in geosynchronous orbits. Two candidates with statistically significant alignments of transients were identified \citep{Villarroel2022c}, of which the more significant candidate happened on 27 July 1952.\footnote{It was incorrectly listed in the original draft as happening on 28 July 1952 due to an error when transforming Julian Date to calendar date, see Villarroel (2024, Debrief) \citep{VillarroelDebrief}.}

In section \ref{sec:Gedankenexperiment}, we show a few such examples based on our toy model, that is introduced in the next section.

\section{The neuro-interface probe as a toy model}\label{sec:toymodel}
\subsection{Arguments for a neuro-interface extraterrestrial probe}

In the following section, we introduce a toy model that includes a neuro-interface extraterrestrial probe capable of communicating with the brain of an advanced organism. We do not claim that this toy model represents actual reality. Rather, we use it in Section \ref{sec:predictions} to demonstrate how a simple model can be used to predict a variety of UAP-related phenomena, ranging from direct physical observations to more subtle effects experienced only by individual observers. The goal is to motivate the design of hypothesis-driven methods in the research of UAP.

We start by first examining the notion of extraterrestrial probes. Probes are devices similar to Voyager or Pioneer, but sent {\em to} the Solar System by an extraterrestrial civilization to collect information. We can further imagine that a civilization might build and send millions or billions of such probes throughout the Galaxy. Bracewell (1960) \cite{Bracewell} suggested that some robotic interstellar probes with the ability to learn and capable of interstellar communication might exist in a giant network of probes. A similar idea was proposed by von Neumann \citep{vonNeumann}, who further suggested that such probes could be self-replicating. In these times of ChatGPT, it is reasonable to assume that each extraterrestrial probe is also equipped with artificial intelligence (AI), which, together with the network of probes, constitutes an extremely powerful Galaxy-wide AI, updating the network within the limitations of the speed of light.

In our toy model, we push the idea even further. We assume the existence of extraterrestrial probes (computers) capable of communicating directly with the human brain, both sending and receiving information in a human-computer interaction. The advantages of connecting to a human brain for collecting information, instead of using optics, are clear. Information from all five senses (sight, hearing, taste, smell, and touch) can be collected simultaneously with the highest possible resolution, with sensitivity to three spatial dimensions and over time.

Even if the latest advances in human-computer interactions presently require installing a chip in the brain (see, e.g., Neuralink\footnote{https://neuralink.com}), we shall avoid assuming that extraterrestrials visit humans at night with the intention to first abduct them and implant a Neuralink chip in their brains.

More simply, we speculatively assume that this extraterrestrial civilization has discovered a method to decode the electrical signals that arise between neurons in the brain and can measure the activity, without any surgical interventions. This could involve measuring the electrical currents inside the brain or the magnetic fields stemming from neural activity, using sensitive sensors operating at a large distance. Alternatively, it could also use quantum processes in microtubules related to consciousness as speculated by Orchestrated Objective Reduction theory \citep{Hameroff} -- or entirely different physical mechanisms.

Instead of assuming that we know anything about the physical mechanisms of such neuro-interface probes, which are based on scientific and technological knowledge far beyond our current capacities, we will introduce a few assumptions about the capabilities of these neuro-interface probes. These probes are:
\begin{itemize}
    \item Capable of collecting information from an individual selected brain.
    \item Capable of collecting information from an individual selected brain at a large distance, ranging from meters to as much as a kilometer or even more.
    \item Capable of recognizing an individual selected human brain among other human brains, and distinguishing it above the noise of other signals.
    \item Capable of communicating back to the individual brain. One possibility is through the use of a directed, narrow electromagnetic beam. Another is with help of quantum communication processes.
\end{itemize}

These probes might be found both inside and outside the atmosphere, as they travel towards the Earth. One possibility is that the neuro-interface extraterrestrial probes are \textit{inside} the atmosphere, while the motherships transporting them to the Solar System are \textit{outside}.  Alternatively, these probes have transmedium capabilities and can travel in and out of the atmosphere.

% :https://webthesis.biblio.polito.it/secure/21994/1/20211215-S277539.pdf
Given that these extraterrestrial probes are equipped with artificial intelligence and can communicate with each other, the probes can share information and update the network's knowledge with anything they learn from humans (or \textit{Drosophila} flies, for that matter). If these probes have been collecting information throughout the Galaxy for billions of years, this network would have an extremely powerful AI capability to make predictions and simulations of events at any location on Earth at a particular time, based on billions of years of data collection regarding events and developments across the Galaxy. Although constrained by factors like quantum uncertainty, thermodynamic irreversibility, and chaos theory \footnote{See Laplace's Demon.}, one could imagine a system that, with significant probability -- though not certainty -- could predict and learn of events such as the flight of a fly across the living room by interfacing with its brain, or the knocking over of a glass by the resident house cat through a similar interface, and could even predict terrorist attacks and natural disasters.

We can only speculate about the fascinating effects that could occur when a human is on the receiving end of such information from the probe.

\subsection{Key predictions}\label{sec:predictions}

What are the key predictions of the existence of such neuro-interface probes? We list here some key predictions.

\begin{enumerate}
    \item One obvious prediction is the presence of physical ET probes either inside the atmosphere, or outside the atmosphere. Optical and infrared images might catch these probes as they reflect light or emit heat, even if they are as small as 10 cm in diameter.
    \item Irrespective of the physical shape of the probe, a human participating in a brain-probe interaction might describe the shape and color of the probe differently from what is seen in the image. In fact, each human might ``see'' the probe entirely differently.
    \item Some probes might be too faint or too small in angular size to be visible to the unaided eye or small mobile cameras at a select distance, and yet visible to the brain as it interacts with the probe, the human witness, and a telescope. 
    \item The presence of humans with prior UAP experiences and subsequent belief near a UAP observatory might increase the likelihood to observe UAP in the sky with external sensors, with an excess on top of the expected bias.
    \item Humans that receive information from the probe might have knowledge, skills or life-altering experiences that are difficult to explain.
    \item Civilian ``UFO hotspots'', i.e. areas with enhanced UFO activity, might show a plethora of abnormalities, from unusual observations to neurological effects in humans and animals.
\end{enumerate}

In case electromagnetic beams are used for communication (as opposed to quantum communication mechanisms), the probe may exhibit a varying electromagnetic spectrum depending on its orientation relative to the observer (anisotropic emission). That would mean that there are angles where the probe can be observed by its own emissivity. This hypothesis might be examined by having two spectrographs separated at a range of a few meters to kilometers from each other.

If quantum communication is used, this opens up the possibility for communication between probes and humans over vast distances, leading to an array of unusual effects related to non-locality, such as instantaneous sharing of information, superposition of sensory perceptions, and observer-dependent phenomena. These effects, while fascinating, are beyond the scope of this paper.

\subsection{Methods of testing it}\label{sec:Gedankenexperiment}

Let us do a simple thought experiment and test the presence of neuro-interface probes. This can be done by either looking for the probes, or by searching for the effects of the probes.

To start with, one could simply look for non-human probes. An academic project doing this is the ExoProbe project \citep{ExoProbe2023}. It aims to build a network of telescopes to search for extra-terrestrial probes near the Earth by searching for transients and unusual streaks. By having multiple telescopes with accurately calibrated separations, we can measure the parallax of any transient that appears due to a short light pulse or a solar reflection from an ET object inside the Solar System. Once more telescopes have been installed and inaugurated, ExoProbe plans to focus on objects with geosynchronous orbits or beyond, where ET objects could stay for millions of years. To carry out such searches, it becomes crucial to remove from the analysis all human-made objects, as there are millions of pieces of human space debris with geosynchronous orbits. 
To achieve this, ExoProbe will employ a method that excludes all human-made satellites and space debris from its search for extraterrestrial probes \citep{Villarroel2024}. Obtaining the spectrum of the object can reveal much about its nature and whether we are dealing with solar reflections or some kind of emission. Getting two spectra from two different sites can also inform us about a potential anisotropy, indicative of what we expect from a neuro-interface probe.

Finding a non-human probe does not necessarily mean it is a neuro-interface probe, even if it has anisotropic emission as predicted. Nevertheless, discovering an extraterrestrial probe, whether or not it has neuro-interface capabilities, would be extraordinary. To investigate the potential for interaction with the human brain, ideally, the extraterrestrial probe should be retrieved. This, of course, requires precise knowledge of the probe's location (or its predictability) and the feasibility of organizing a space mission to recover it. Assuming a successful probe retrieval mission and that the probe is brought to a laboratory, experiments could then commence to determine if it has any effects on humans in proximity. This direct approach would be optimal: astronomers would be tasked with detecting and identifying the probes with an experiment similar to ExoProbe, a space company to bring the probe down to the Earth, and neuroscientists and psychologists would be tasked with testing whether the probe interacts with the human mind or not.\footnote{We are astronomers and do not pretend to understand humans. We hope scientists from other disciplines will help us with it.}

Nevertheless, it is more likely that complementary methods will be needed to explore the potential human-probe connection. An alternative method for detecting neuro-interface probes involves testing the cognitive abilities of a UFO witness at the time of reporting. It is conceivable that a neuro-interface probe might transmit information to the witness in a manner that could be mistaken for an enhancement in cognitive abilities, or even more remarkably, as signs of extrasensory perception. The psychological literature contains numerous examples of experiments that have purportedly measured such effects, some of which could be adapted for inclusion in the UFO reporting process. This however  requires a different expertise than astronomers can assist with.

Finally, UFO 'hotspots' could be equipped with observatories or dedicated UFO observation stations to determine if probe detection is more likely in these areas. Given the claims of localized effects, it is reasonable to assume that these hotspots may be particularly valuable for searches within the atmosphere. For UFO researchers interested in investigating atmospheric anomalies, these hotspots offer an excellent opportunity to observe potential effects on animals and humans in the vicinity, as well as to document any UFO sightings. 

We can only hope that these hotspots have a civilian \textbf{or NHI origin} and are not covert military testing grounds.

\section{Ethical considerations in UAP research}\label{sec:ethics}

Research on UAP does not occur in isolation; it involves instruments, human witnesses, national security interests, geopolitical tensions, and purely scientific interests. Misidentified UAP can potentially trigger nuclear conflicts  \citep{Sol2024a}, placing an additional burden on civilian astronomers that research on distant stars and planets does not. In this section we will consider the ethical side effects of conducting UAP research. We will not discuss the dissemination of results but will focus on how the chosen methodology might impact society.

A case that requires consideration involves actual UFO reports, which are reports made by witnesses to an UFO organization. Such reports are often collected but not published due to the sensitive information. A comprehensive UFO report includes observational information on a sighting that may be of earthly or other-worldly origin, as well as the name of the witness, personal information, conditions of the testimony, and the impact on the witness. Some reports may involve sensitive government activities, such as military tests, and this information is best kept within national borders.

While a single UFO report might be linked to one event, thousands of reports can reveal patterns, such as military test sites. Therefore, UFO reports are sensitive from a national security perspective.

From a personal data perspective, witnesses might be wary of who is allowed to read the file and identify the witness. In the European Union, GDPR rules protect the trading of personal data across continents, although not every UFO organization adheres to these rules. Many report centers aim to collect as much data as possible and might reach out to organizations outside their country's borders. This raises the question of how scientists and amateur organizations can develop data handling routines for UFO data, considering both personal data and potentially sensitive data from a national security perspective. 

A simple ethical rule might be: do not trade personal UFO reports across national borders and avoid using UFO report applications developed by other countries.

A second issue arises from the scientific search of UAP. The closer to the ground one searches for unknown flying objects, the greater the likelihood that the unknown object is human-made. Observing everything that moves in the sky and categorizing all flying objects while waiting for the ``unknown'' inevitably means categorizing all military vehicles as well. Using simple reasoning, imagine a UAP research group using advanced instrumentation to collect data on all moving objects in the air, both natural and artificial. In their search for the ``unknown'' object, they must first categorize every flying object. This requires carefully identifying military jets, birds, and Boeing 434s. However, in the process, they inevitably gather information on the number and types of military vehicles passing through that particular airspace. This inadvertently informs them about the country's air force, regardless of their original intent. Now, if the collaboration is international, this data might be shared with other countries, including those outside NATO. It is not difficult to see how such an open data-gathering process can become sensitive. A quest to search for UAP in an open manner might unintentionally survey the sky in ways that are unappreciated by the military.

This responsibility is unnecessarily large, especially in international collaborations. Learning about military interests does not teach civilian astronomers about the mysteries of the universe but solely about human technology. Civilian scientists should not be concerned with whether UAP are secret experiments by other countries, but rather, are they non-human intelligence that can increase our knowledge of the universe? One way to avoid this issue is by focusing on observations outside the atmosphere and in relatively unpopulated orbits around the Earth. Another method is to focus on clear hypotheses related to UAP research and searching for specific, interesting signatures, rather than categorizing every \textit{Larus marinus} or airplane. A simple rule is: whatever decreases the number of false positives also decreases the number of objects we should not encounter.

A third issue is related to claimed ``hotspots.'' Famous public hotspots include supposedly Skinwalker Ranch\footnote{A tv-series aired at the \textit{History Channel}, follows UFO sightings at the Skinwalker Ranch, a ranch owned by Brandon Fugal.} and Hessdalen \citep[e.g.][]{Teodorani2004,Hessdalen}.\footnote{Project Hessdalen: www.hessdalen.org} It is commonly claimed that UFOs appear near nuclear sites and military facilities \citep[e.g.][]{Hastings2015}. However, many countries protect imagery of these sites by law, and inspecting images taken by satellites from other countries might still be illegal. Regardless of whether there are more flying saucers near nuclear power plants or not, we recommend not searching for aliens near nuclear sites or other military installations.

\section{Conclusions}

The field of Unidentified Anomalous Phenomena (UAP) is a new, emerging multidisciplinary area of research where the expertise of observational astronomers is essential. Astronomers have observed UAP for hundreds of years, and even in systematic studies such as Project Moonwatch in the late 1950s, UAP were reported, contrary to many beliefs.

Unfortunately, scientific methods to study UAP are often too broad and insufficiently focused on testing the most relevant hypotheses. We propose that astronomers abandon the attempts to understand the origin of UAP at large, and instead concentrate on examining the ET/NHI hypothesis for UAP using the scientific method. We specifically propose searches for neuro-interface NHI probes. With a neuro-interface probe toy model, the gap between human experience and observations can be reconciled, and careful experiments can be more easily designed, rather than relying on UFO witness reports as has been done over the past 80 years. The UAP question is as scientific as any other problem, and it is the work of astronomers to find simple and effective ways to decode this enigma.

\section{Acknowledgments}
\label{sec:ackn}
We thank Geoff Marcy for useful discussions and suggestions, in particular with the historical examples in astronomy. A heartfelt thanks goes also to Karl Nell, for detailed and deep feedback on the manuscript with many great suggestions. We also thank Rudolf Baer, Stephen Bruehl, Baptiste Friscourt,  Richard Geldreich and Rex Groves, Joel Jewitt, Eugene Jhong, Francisco Ricardo, Mark Rodeghier for helpful suggestions. ExoProbe is supported by an anonymous donor, to whom we are deeply grateful.


%%\bibliography{sample63}{}
{99}

%%\bibliography{references.bib}

\end{document}